\documentstyle[preprint,eqsecnum,aps,prb]{revtex}
\begin{document}
\title{Investigation of laser ablated ZnO thin films grown with Zn metal target: a
structural study.}
\author{A. Fouchet, W. Prellier\thanks{%
prellier@ismra.fr} and B. Mercey}
\address{Laboratoire CRISMAT, CNRS\ UMR 6508, 6 Bd du Mar\'{e}chal Juin, \\
F-14050 Caen Cedex, FRANCE.}
\author{L. M\'{e}chin}
\address{Laboratoire GREYC, CNRS\ UMR 6072, ENSICAEN\ et Universit\'{e} de Caen,\\
6 Bd du Mar\'{e}chal Juin, F-14050 Caen Cedex, FRANCE.}
\author{V. N. Kulkarni\thanks{%
On leave from Indian Institute of Technology, Kanpur, India.} and T.
Venkatesan}
\address{Center for Superconductivity Research, University of Maryland,\\
College Park, Maryland 20742-4111, USA}
\date{\today}
\maketitle

\begin{abstract}
High quality ZnO thin films were gown using the pulsed laser deposition
technique on (0001) Al$_2$O$_3$ substrates in an oxidizing atmosphere, using
a Zn metallic target. We varied the growth conditions such as the deposition
temperature and the oxygen pressure. First, using a battery of techniques
such as x-rays diffraction, Rutherford Backscattering spectroscopy and
atomic force microscopy, we evaluated the structural quality, the stress and
the degree of epitaxy of the films. Second, the relations between the
deposition conditions and the structural properties, that are directly
related to the nature of the thin films, are discussed qualitatively.
Finally, a number of issues on how to get good-quality ZnO films are
addressed.
\end{abstract}

\newpage

\section{Introduction}

Zinc oxide (ZnO) is a II-VI semiconductor highly transparent (85-95\%) in
the visible region with a wide and direct band-gap of about 3.3eV at room
temperature and a high exciton binding energy of 60 meV\cite{zno0}. This
semiconductor is very attractive because it has many applications such as
transparent conductive contacts, solar cells, laser diodes, ultraviolet
lasers, thin films transistors and others\cite
{zno1,zno2,zno3,zno4,zno5,zno6,zno7}.

ZnO\ is an oxide that can be grown, as a thin film, by many deposition
techniques including chemical vapor deposition, radio frequency sputtering,
magnetron sputtering, sol-gel, ion-beam assisted, molecular beam epitaxy and
pulsed laser deposition\cite
{Tech1,Tech2,Tech2b,Tech2c,Tech3,Tech4,Tech4b,Tech5}. The zinc oxide shows a 
$n$-type conduction, without intentional doping, and crystallizes in a
wurtzite structure that contains large voids that can easily accommodate
interstitial atoms. The intrinsic point defects i.e. zinc interstitials have
been proposed to be the donors causing $n$-type conduction\cite{Look}.
However, in order to control the doping content, it is necessary to remove
oxygen vacancies (since the lack of oxygen and an excess of zinc leads to a $%
n$-type semiconductor). For this reason, the pulsed laser deposition (PLD)
technique is interesting because it gives the advantage of carrying out the
growth in a high-oxygen partial pressure (typically around 0.2 torr)\cite
{Tech5}. Usually, the films are obtained from a ZnO target\cite
{Tech5,Kawa,PerriereFemto} but we have recently developed a variant
technique in which the oxide films are grown from a metal target\cite
{Mike,Arnaud}. We have also used this technique to synthesis ZnO\ films
described in this article.

Fabricating high quality epitaxial thin films with a perfect $c$-axis
texture, a low density of defects (i.e. low strained films) and an extremely
smooth surfaces, are important for many applications.\ Thus, it is necessary
to understand the growth mechanisms of ZnO films that influence the
structural and the optical properties. Here, we report the growth of ZnO
films on sapphire (0001)-oriented by PLD, using a Zn metal target. A number
of issues on how to control the growth conditions in order to get
good-quality ZnO films are addressed in this paper. In particular, the
effects of the deposition parameters on the degree of epitaxy and defects
density of laser-ablated ZnO films, using a Zn metal target, are
qualitatively discussed.

\section{Experimental}

The ZnO films were grown utilizing the pulsed laser deposition technique
(Lambda Physik, KrF laser $\lambda $=248nm) \cite{PLD} by firing a zinc
metal target (99.995\%) as purchased (NEYCO, France) and without further
preparations. The films are deposited on (0001) Al$_2$O$_3$ substrates at a
constant temperature under a flux of pure oxygen. These two parameters
(temperature and oxygen pressure ) were systematically varied. The substrate
temperature ranged from 400 to 750${{}^{\circ }}C$ and the pressure was
changed from 0.03 to 0.2 torr of O$_2$. Thus, different series of films are
obtained. At the end of the deposition, the films were slowly cooled to room
temperature (20${{}^{\circ }}C/\min $) under 225 torr of oxygen.

The structural study was done by X-Ray diffraction (XRD) using a Seifert XRD
3000P for the $\Theta -2\Theta $ scans and the $\omega $-scan (tilting). An
X'Pert Phillips was also employed for the in-plane measurements ($\Phi $%
-scans). Both diffractometers utilized the Cu, K$\alpha 1$ radiation ($%
\lambda =0.15406nm$). In order to obtain additional information on the
structural properties of the ZnO films, we have determined the thin film
strains. This technique used the distance between the atomic planes of a
crystalline specimen as an internal strain gage\cite{Strains}. The plane
spacing $d_{hkl}$ is normal to the diffraction vector $\overrightarrow{L}$.
One can define a strain $\varepsilon $ , along this diffraction vector $%
d_{hkl}$, $\varepsilon =(d_{hkl}-d_0)/d_0$ where $d_0$ is the unstressed
plane spacing of the ($hkl$) planes (i.e. the value of the bulk). To measure
the stress, we used the sin$^2\Psi $ technique \cite{Strains}. Briefly, in
this model, the strain is defined as follows: $\varepsilon =A\sin ^2\Psi $+$B
$ (where $A$ and $B$ are constants that depend on the strain along the
surface direction, the Young's modulus, Poisson's ratio and the stress along
the direction, for details see Ref.15). Additional informations are given
from the XRD measurements by calculating the grain size of the crystallites
along the $c$-axis using the Scherrer's formula\cite{Sherrer} :

$\beta _{\frac 12}=\frac{0.9\lambda }{D\times \cos \theta }$ ~ (1)

where $\lambda $ is the X-ray wavelength, $\beta _{\frac 12}$ is the (0002)
peak width (in radian), $D$ the crystal size and $\theta $ the Bragg
diffraction angle.

Rutherford back scattering spectroscopy (RBS)/ion channeling using a
well-collimated 1.5-3 MeV He$^{+}$ ions has been used to determine the
epitaxy quality and the film thickness. Surface morphology and roughness
measurements were determined by Atomic Force Microscopy (AFM) in tapping
mode using a Multimode Nanoscope III (Digitals Instruments).

Optical measurements were investigated with an UV-Visible spectrophotometer
(Cary 100 scan) to determine the energy band gap of the different films. In
this method, the fundamental absorption, which corresponds to electron
excitation from the valence band to conduction band, is used to determine
the value of the optical band gap through the following relation:

$\ \ (h\nu \alpha ){{}^2}=C(h\nu -E_g)$\ (2)

where $~\alpha $ is the absorption coefficient, $C$ is a constant, $h\nu $
is the incident photon energy and $E_g$ the band gap.

\section{Results}

\subsection{Study as a function of the deposition temperature}

A series of films grown in 0.1 torr of O$_2$ were investigated. The
resulting XRD patterns of ZnO exhibit two diffraction peaks observed around 2%
$\Theta $ close to 34.48${{}^{\circ }}$ and 72.66${{}^{\circ }}$. They are
characteristic of the hexagonal ZnO wurtzite, the $c$-axis being
perpendicular to the substrate plane\cite
{ceramtarget,ceramtarget1,ceramtarget2}. The out-of-plane lattice parameter
is calculated to be close to 5.205 A\ which corresponds to the theoretical
bulk value\cite{bulkvalue}. Moreover, the out-of-plane lattice parameter is
almost constant, equal to 5.198$\pm $0.004 A\ in the temperature range
500-700${{}^{\circ }}C$, which is slightly shorten, as compared to the bulk,
but in agreement with a tensile stress within the plane (see below in the
text). Below 500${{}^{\circ }}C$ and above 700${{}^{\circ }}C$, the
out-of-plane lattice parameter is much lower (5.185 A). Nevertheless, the
full-with-at-half-maximum (FWHM) of the rocking-curve $\omega $-scan
measured around the 0002 reflection (tilting) is decreasing from 1.8${%
{}^{\circ }}$ at 400${{}^{\circ }}$C to 0.3${{}^{\circ }}$ at 600${{}^{\circ
}}C$. It remains constant from 600${{}^{\circ }}C$ to 750${{}^{\circ }}C$
(see Fig.1) The in-plane epitaxy was investigated by measuring the $\Phi $%
-scan (recorded around the 2$\overline{11}3$ reflection) at different
deposition temperatures. The resulting plot is given in Fig.3.\ While the
films grown above 600${{}^{\circ }}C$ exhibit 6 diffraction peaks separated
of 60${{}^{\circ }}$ in agreement of the six-fold symmetry of the hexagonal
lattice of the ZnO, the films deposited below display 12 diffraction peaks
separated of 30${{}^{\circ }}$. In fact, these 12 peaks correspond most
probably to two different series of 6 peaks i.e. two series of crystallites
(or oriented domains) that are epitaxially different. The FWHM of the $\Phi $%
-scan (twisting) is also indicated on the left part of Fig.3. It confirms
the degradation of the epitaxy as the deposition temperature is decreasing
(see also Fig.1).

In order to evaluate the stress of the ZnO films, we have used the sin${{}^2}%
\Psi $ model\cite{Strains} that allows to determine the residual stress in
epitaxial plane. The stress (determined as the slope of the curve when
plotting the strain $\varepsilon $ vs. sin${{}^2}\Psi $) is decreasing when
increasing the substrate temperature.\ The residual stress is calculated to
be 107 MPa at 600${{}^{\circ }}C$ and 100.5 MPa at 700${{}^{\circ }}$C. Note
that the decrease of the residual stress when increasing the deposition
temperature is in a perfect agreement with the decrease of the FWHM of both
the (0002) and the (2$\overline{11}3$) reflections (see Fig.1). In addition,
this value is much lower than previously reported ones (330-500 MPa)\cite
{PerriereFemto} but these films were grown from a ZnO\ target and the
deposition performed at 550${{}^{\circ }}C$ under 3.75$\times $10$^{-4}$
torr.

The absorbance measurements investigated between 200 and 900 nm show that
the films are highly transparent in the visible region and present a steep
fall off around 380 nm. This fundamental absorption can be used to determine
the value of the optical band gap. Its dependence on the deposition
temperature, is given in Fig. 2a. The band gap energy is increasing with the
temperature as a result of an increase of the number of defects \cite{GAP}.
However, between 550-700${{}^{\circ }}C$, the energy gap is almost constant
and close to $3.20$ eV, a characteristic of high quality ZnO films. The
grain size has also been calculated using the Sherrer's formula. The grain
size increases with the temperature in the range of 600${}${-700 }${^{\circ }%
}C$ up to 25 nm (see Fig.2). As a result, when the grain size is increasing,
the epitaxy of the films is improving and the energy band gap is increasing
to a constant value.

These experiments suggest that at a pressure of 0.1 torr of O$_2$, the
deposition temperature should be around 600${{}^{\circ }}C$. Thus, in a
second step, we have investigated the influence of the oxygen pressure for a
fixed temperature of 600${{}^{\circ }}C$.

\subsection{Study as a function of the oxygen pressure}

Basically, the XRD of the films ($\Theta -2\Theta $), grown at various
oxygen pressure, resemble the previous one. Two diffraction peaks
corresponding to $0002$ and the $0004$ reflections are predominant and
indicate that the film has a $c$-axis orientation with an out-of-plane
lattice parameter around 5.2 A. In Fig.2b, the grain size and the band gap
exhibits a maximum indicating an optimum value of the oxygen pressure close
to 0.1 torr\cite{Chen04}. In the same way, the FWHM\ of rocking curve
(tilting) presents a minimum (0.3${{}^{\circ }}$) for a pressure of 0.1 torr
(see Fig.1).\ Similar behavior is observed when plotting the FWHM\ of the $%
\Phi $-scan (twisting): the minimum value (0.9${{}^{\circ }}$) is also seen
at a pressure close to 0.1 torr of O$_2$ (see Fig.1). Such values are
comparable to the results of other groups which used ceramic targets\cite
{ceramtarget,ceramtarget1,ceramtarget2}. These results also confirm that a
pressure of 0.1 torr will give the better epitaxy (as seen from $\Phi $%
-scan), better texture (see the values of the FWHM) and larger
nano-crystallites (30 nm using Sherrer's formula) of the ZnO films. The low
value obtained at 0.2 torr for the twisting angle will be explained latter.
We have also noted that the measured value of the out-of plane lattice
parameter is 5.1982 A, which is lower than the bulk value (5.2066 A\cite
{bulkvalue}). This indicates a compression ( $\sim $0.1\%) of the
out-of-plane parameter and an expansion of the in-plane lattice parameter
(3.253 A\ in the film to be compared with 3.2498 A in bulk\cite{bulkvalue})
as observed for many oxide films\cite{StrainW,StrainW1}. The tensile stress
within the plane of the substrate is confirming by the positive slope of the
curve (strain $\varepsilon $ versus sin${{}^2}$$\Psi $). The resulting data
are given in Table 1 for a series of films. At 0.1 torr, the residual stress
is minimum (107 MPa) as compared to the films grown at 0.2 torr (334 MPa) or
0.03 torr (232 MPa). In addition, the film prepared at 0.1 torr approached
the $c$-axis value of the bulk\cite{bulkvalue} in agreement with a low
residual compressive stress. This is also in perfect accordance with the
evolution of the FWHM along both $\omega $ (tilting) and $\Psi $ (twisting)
scans \smallskip (Fig.1) where the minimum values are obtained for a
pressure of 0.1 torr.

The structural analysis was completed using the measurements of ion
channeling using RBS and the results spectra (random and aligned) recorded
for three different oxygen pressures are given in Fig.4. The $\chi _{\min }$
value, which indicates the quality of the film is given a minimum at 0.1
torr (Aligned yield of 8.7\%) as compared to the film grown at 0.03 torr
(aligned yield: 24\%). Surprisingly the film grown at 0.2 torr is not
epitaxied (no alignment). The low experimental $\chi _{\min }$ value
indicates a very low level of structural defects in this film. On the
contrary, a high $\chi _{\min }$ value is correlated to a presence of large
amount of defects in the crystallites. The presence of defects leads to a
broadening of the diffraction peaks as observed in the plot of the FWHM\
vs.\ oxygen pressure (Fig.3). Note also, that the thickness of the film was
also determined with the simulation of the structure and we found that it is
increasing with the increase of the oxygen pressure.

The morphology and the surface roughness, determined by AFM\ measurements,
are represented in Fig.5. The roughness seems to decrease with the increase
of oxygen concentration and in the same time, the grains size on the surface
is changing. For the film grown at low pressure (0.07\ torr), the grains
size is around 250 nm and seems relatively constant. Furthermore the
roughness of this film is about 50 nm. On the contrary, the film grown at
higher pressure (0.1 torr) shows two types of grains with two different
sizes: higher 500 nm and smaller 160 nm than for the film at low pressure.
The roughness is around 50 nm. Finally for the film grown at 0.2 torr, the
grains size is larger and seems inhomogeneous, but the roughness (rms) being
of only 23 nm.

These studies were followed by the determination of the energy band gap,
plotted in Fig.2b versus the pressure of oxygen. The film deposited under
0.1 torr presents the widest energy band gap with the value of 3.20 eV.
Furthermore the comparison of the gap with the grain size shows the same
tendency than with the temperature: when the grains size is larger, the
energy band gap between the conduction band and valence band is increased.

\subsection{Discussion}

These experiments performed on a series of ZnO\ thin films clearly show that
there is a correlation between the deposition parameters (substrate
temperature, oxygen pressure) and many structural parameters including
crystallinity, grain size, epitaxy in-the-plane of the substrate, texture
along the out-of-plane direction, residual stress, roughness and defects. As
observed from the experiments, the best compromise for formation of ZnO
films is obtained with a deposition temperature of 600 ${{}^{\circ }}C$ and
an oxygen pressure of 0.1 torr. In this section, we will discuss the effects
of the pressure and the temperature upon the formation of ZnO. To understand
the influence of these parameters, two conditions must be taken into
account: the dynamics of the ablation materials and the nucleation\cite
{Chrisey}.

At the beginning, the zinc target, is in the metal state and the vapor
species ablated out are constituted by zinc element only. The ZnO formation
is due to the reaction with the ambient gas of the chamber. The collisions
between the ejected species and the oxygen increases with the oxygen
pressure. On one hand, if the pressure of oxygen is low, the ion fraction
and the kinetic energy of the evaporated target material are not really
reduced. Furthermore the oxidation could not be completed, allowing some
defects in the film and some interstitial zinc. This number of defects is
confirmed with the (2$\overline{11}$3) $\Phi $-scan of the film grown at
0.03 torr which shows 12 peaks separated by 30${{}^{\circ }}$ and confirms
the low epitaxial quality. On the other hand, if the quantity of oxygen is
too high, the vapor species can undergo enough collisions that nucleation
and growth of these vapor species form particulates before their arrival at
the substrate. In order to understand the number of collisions, the mean
free path is approximately 5 cm for a pressure of 10$^{-3}$ torr and 0.05 cm
at a higher pressure of 0.1 torr. Thus, a high ambient pressure increases
the size of ultrafine particulates (which are deposited in the film) and the
thickness layer, and creates a large number of defect.

Furthermore, the nucleation process depends on the interfacial energies
between the three existing phases: substrate, the condensing material and
the vapor. The minimum-energy shape of a nucleus is like a cap. The critical
size of the nucleus depending on the driving force, $i.e.$, the substrate
temperature (fixed: first part of the study) and the deposition rate. A
characteristic of small supersaturation (0.03 torr) for large nuclei is the
creation of isolated patches (islands) of the film onto the substrate which
subsequently grow and coalesce together. As the supersaturation increases,
the critical nucleus shrinks until its height reaches on atomic diameter,
resulting in a two-dimensional layer (0.1 torr). For large supersaturation,
the PLD process would cause rapid nucleation of very small clusters. This is
confirmed with the calculation of grain size (see Fig.2b) which indicates
small crystallites: 18 nm and 20.5 nm along the $c$-axis for the films grown
at 0.03 torr and 0.2 torr, respectively, compared with the films grown at
0.07 torr and 0.1 torr (around 30 nm). Furthermore, the crystalline film
growth depends on the surface mobility of the adatom (vapor atoms).
Normally, the adatoms will diffuse through several atomic distances before
sticking to a stable position within the newly formed film. The surface
temperature of the substrate determines the adatom's surface diffusion
ability and a full oxidation of the zinc atom. The high temperature (see
first part) favors rapid and defect-free crystal growth; whereas, low
temperature (between 400 and 450 ${{}^{\circ }}$C) or large supersaturation
crystal growth may be overwhelmed by energetic particle impingement,
resulting in disordered or even poor crystallized structures.

~Finally, the dependance of the number of defects from the gap calculation
can be observed with optical measurements. The two films deposited at 0.2
torr (no epitaxy) and 0.03 torr (oriented domains) show very low energy band
gaps. Thus, optical results follow the same trend as the structural analysis.

\section{Conclusion}

High quality ZnO thin films were grown using the pulsed laser deposition
technique on (0001) Al$_2$O$_3$ substrates starting from a Zn metal target.
The growth was performed under high oxygen pressure (0.03-.2 torr) in the
temperature range 400-750${{}^{\circ }}C$. We have studied the influence of
the growth conditions (deposition temperature, oxygen pressure) upon the
structural properties that are directly related to the nature of a thin
films: out-of-plane alignments, in-plane epitaxy, grain size, distribution
of the stress in the different directions etc...

We concluded that the optimum deposition conditions leading to high quality
ZnO\ thin films with good optical properties when using a Zn metal target
are 600${{}^{\circ }}C$ and 0.1 torr of O$_2$. This new alternative method
of using a Zn metal target, as compared to ceramic target, allows a better
control of the zinc oxidation and lead to good structural properties that
can be used to many interesting oxides such as VO$_2$, WO$_3$ or TiO$_2$.

Acknowledgments

W.P. acknowledges the Centre Franco-Indien pour la Promotion de la Recherche
Avanc\'{e}e/Indo-French Centre for the Promotion of Advance Research
(CEFIPRA/IFCPAR) for its partial financial support (Project N${{}^{\circ }}$%
2808-1). A.F. also thanks the ''Centre National de la Recherche
Scientifique'' (CNRS) and the ''Conseil Regional de Basse Normandie'' for
his BDI fellowship.

We also thank Dr.\ H.\ Eng for careful reading of the article.

\newpage Figures Captions:

Figure 1: FWHM\ of the $\omega $-scan (recorded around (0002) reflection,
see left axis) and $\Phi $-scan (recorded around the (2$\overline{11}$3)
reflection, see right axis) plotted as a function of the oxygen pressure
(for a deposition temperature of 600${{}^{\circ }}C$, see bottom axis) and
as a function of the deposition temperature (for an oxygen pressure of 0.1
torr, see top axis).

Figure 2: Energy band gap (square) and crystal size (circle) as a function
(a) of the deposition temperature (the oxygen pressure is fixed at 0.1 torr)
and (b) of the oxygen pressure (the temperature is 600 ${{}^{\circ }}C$).
see text for details.

Figure 3: $\Phi $-scan recorded around the (2$\overline{11}$3) diffraction
peaks for different ZnO films grown under 0.1 Torr of O$_2$ at various
substrate temperatures. The FWHM\ of the peaks in the $\Phi $-scan is
indicated in the left panel of the graph.

Figure 4: Rutherford backscattering random (open circle) and aligned (full
square) for a series of films grown under different oxygen pressures at 600${%
{}^{\circ }}C$. The aligned yield and the thickness of the films are given.
The simulation of the structure is also indicated (black line).

Figure 5: AFM images of ZnO\ films deposited at 600 ${{}^{\circ }}C$ with
(a), (b), (c) the height of the film grown under 0.2 torr, 0.1 torr and 0.03
torr of O$_2$, respectively and (d), (e), (f) the amplitude signal recorded
in the same scans.

\smallskip \newpage

Tables:

Table 1: Evolution of various structural parameters for a series of films.

\end{document}